\documentclass[a4paper,11pt]{article}
\pdfoutput=1 
\usepackage{jcappub}
\usepackage[T1]{fontenc}
\usepackage{color}
\usepackage{graphicx}
\usepackage{float}
\usepackage{amsmath,amsfonts,amssymb}
\usepackage[utf8]{inputenc}
\usepackage{graphicx}
\usepackage{epsfig}
\usepackage{subfigure}
\usepackage{color}
\usepackage{comment}
\usepackage{slashed}
\usepackage{yfonts}

\title{\boldmath A simple $F({\cal R},\phi)$ deformation of Starobinsky inflationary model}
\author[a,b]{Dhimiter D. Canko,}
\author[a]{Ioannis D. Gialamas}
\author[a]{and George P. Kodaxis}
\affiliation[a]{Department of Physics, University of Athens,\\Zographou 15784, Greece}
\affiliation[b]{Institute of Nuclear and Particle Physics, NCSR "Demokritos",\\ Agia Paraskevi 15310, Greece}
\emailAdd{jimcanko@phys.uoa.gr}
\emailAdd{I.Gialamas@phys.uoa.gr}
\emailAdd{g.kodaxis@hotmail.com}

\abstract{We study  a model including a real scalar field $\phi$  non-minimally coupled to $F({\cal R})$ gravity, which is conformally equivalent to an Einstein-Hilbert theory, involving two real scalar fields. We consider three special cases of the potential of the field $\phi$ in the $F({\cal R})$-frame: a vanishing potential, a mass term and a Higgs potential. All these lead to non-trivial two-field potentials in the Einstein-frame which in particular directions resemble the well-known Starobinsky model. We find, that all these cases can yield viable inflationary models in complete agreement with current observational data.}

\begin{document}
\maketitle
\flushbottom


\section{Introduction} \label{Introduction}
\label{sec:intro}

The mechanism of cosmological inflation \cite{Staro,Sato:1980yn,Guth,Linde:1982,Lyth:1998xn} was first introduced in 1980's in order to solve crucial problems of the Big Bang Cosmology, such as the horizon, the flatness and the Monopole problems. Inflation is a period of accelerated (quasi-de Sitter) expansion of the very early Universe, which elegantly allows for near large-scale homogeneity and spatial flatness of our Universe. An extra bonus of introducing inflation in Standard Cosmology is that it can explain the formation of large-scale structure, being the only known mechanism to do this. Quantum fluctuations during the inflationary epoch presumably seeded the perturbations which grew under gravitational instability into the structures we observe today \cite{Mukha}. Due to this, the inflationary mechanism has been intensively studied resulting to a better theoretical understanding of it. Also, in recent years, the interest in inflationary cosmology has grown considerably because of the great amount of data made available sourcing from various cosmological surveys. Despite its successful predictions the origin of inflation is not well understood, as yet. It is more like a phenomenological construction, whose origin should be sought in some fundamental theory, such as high-energy particle physics or gravity.

In particle physics, superstring theory and supergravity, multiple scalar fields are involved and some of them may play the role of inflaton. Furthermore, in curved spacetime and in the context of renormalization of scalar fields we have the arise of non-minimal couplings between scalar fields and the Ricci scalar \cite{Birrell,Birrell2}. Thus it is reasonable to search for inflationary models which include many fields non-minimally coupled to gravity, whose potential energy dominates the energy-momentum tensor and drives inflation \cite{Wands2007,Sfakianakis,Sfakianakis2,Kaiser,Riotto,Peterson,Peterson2}. For a model of inflation to be viable it should be in agreement with the recent observational constraints for the spectral index, the tensor-to-scalar ratio and non-gaussianity (for multi-field inflation). These quantinties are defined in the context of cosmological perturbations \cite{Starobinsky:1979ty,Starobinsky:1982ee,Guth:1982ec,Starobinsky:1986fxa,Mukha,Gordon:2000hv,Starobinsky:2001xq,Basset,Malik,Lalak:2007vi,Langlois:2008mn,White:2012ya,Qiu:2014apa} and their constraints obtained from observetions of cosmic microwave background (CMB), according to \cite{Planck,Planck2}, are: 
\begin{equation}\label{obs} 
\begin{aligned}
n_s&=0.9649 \pm 0.0042 \hspace{1.48cm}\text{68 $ \% $ CL}\, ,
\\  r&<0.064  \hspace{3.24cm} \text{95 $ \% $ CL}\, , \\
f_{NL}&=0.8 \pm 5.0 \hspace{2.67cm} \text{68 $\%$ CL} \, .
\end{aligned}
\end{equation} 

Among the single-field models of inflation, various classes of models can be in agreement with the aforementioned constraints. The first is the well studied non-minimally coupled Higgs inflation \cite{Bezrukov:2007ep,Bezrukov:2008ej,Barvinsky:2008ia,Barvinsky:2009fy,DeSimone:2008ei,Barbon:2009ya,Lerner:2010mq,Barvinsky:2009ii,Bezrukov:2013fka,Kamada:2012se,Ketov:2012jt,Bezrukov:2010jz,Hamada:2013mya,Bezrukov:2014bra,Allison:2013uaa,Salvio:2015kka,Hamada:2014wna,Calmet:2016fsr,Rubio:2018ogq,Enckell:2018kkc} which provides a particle origin to inflaton, but is also strongly connected with gravity. The second is the class of models of chaotic inflation \cite{Linde:1983gd,Linde:1985ub,Linde:2014nna} and its variants, and another is natural inflation models \cite{Freese:1990rb,Adams:1992bn}. Also a well-known motivated model is the Starobinsky model of inflation \cite{Staro}, which remarkably was proposed almost four decades ago and furnishes a gravitational origin to inflaton. This model can be seen as the simplest inflationary model within the context of $\cal F(R)$ theories of gravity \cite{Capozziello:2006dj,Briscese:2006xu,Sotiriou:2008rp,Appleby:2009uf,Ishak:2018his,Nojiri:2008nt,DeFelice:2010aj,Capozziello:2011et,Clifton:2011jh,Rinaldi:2014gua,Bamba:2014wda,Odintsov:2016vzz,Chakraborty:2016ydo}, as the only extension from Einstein Gravity is the addition to the Hilbert - Einstein action of an extra ${\cal R}^2$ term
\begin{equation}
\label{star}
S=\int d^4x \, \sqrt{-\tilde{g}} \left[-\frac{1}{2}{\cal \tilde{R}} + \frac{1}{12M^2}{\cal \tilde{R}}^2 \right] \, ,
\end{equation} 
where above $M$ is a parameter with dimensions of mass, $\tilde{g}_{\mu\nu}$ is the metric tensor and $\cal \tilde{R}$ is the Ricci scalar. In this paper, the reduced mass Planck is dimensionless and equal to 1, and we use the $(+,-,-,-)$ spacetime signature notation. This action is classically equivalent, through a conformal transformation, to the following scalar-tensor theory with a non-minimal coupling between the scalar field $\phi$ and gravity
\begin{equation}
\label{stara}
\begin{aligned}
S =&\int d^4x \, \sqrt{-g} \left[-\frac{1}{2}{\cal R}+ \frac{1}{2}g^{\mu\nu}\partial_{\mu}\phi\partial_{\nu}\phi - \frac{3}{4}M^2 \left(1-e^{-\sqrt{\frac{2}{3}}\phi} \right)^2 \right] \, .
\end{aligned}
\end{equation} 
Using the slow-roll inflation formalism and the agreement with the observables the value of $M$ is restricted to $M\simeq1.3 \cdot 10^{-5}$.

The great success of Starobinsky inflation model and its elegant physical interpretation of inflaton has led to an intensive study of inflationary models that are extensions or modifications of this model in the framework of $F({\cal R})$ theories or $F({\cal R}, \phi)$ theories studied in the metric formalism \cite{Berkin:1990nu,Huang:2013hsb,Ben-Dayan:2014isa,Broy:2014xwa,Artymowski:2015mva,Asaka:2015vza,Sebastiani:2015kfa,Elizalde:2018now,Elizalde:2018rmz,Rador:2007wq,Kannike:2015apa,Ketov,vandeBruck:2015xpa,Wang:2017fuy,Liu:2018hno,Castellanos:2018dub,Gundhi:2018wyz,Gorbunov:2018llf,He:2018gyf,He:2018mgb,Pi:2017gih,Kohri,Ema:2017rqn,delaCruz-Dombriz:2016bjj,Karamitsos:2017elm,Chakraborty:2018scm,Karam:2018mft,Kubo:2018kho} (see for instance \cite{Myrzakulov:2015qaa} for a beautiful discussion of inflation in the Jordan frame). The latter have the meaning of being $F({\cal R})$ theories in the presence of scalar fields, which are in general non-minimally coupled to gravity. An alternative variational principle leading to the equations of motion of General Relativity, which has been intensively studied for
cosmological purposes \cite{Tamanini:2010uq,Bauer:2010jg,Enqvist:2011qm,Borowiec:2011wd,Stachowski:2016zio,Rasanen:2017ivk,Tenkanen:2017jih,Racioppi:2017spw,Markkanen:2017tun,Fu:2017iqg,Enckell:2018hmo,Antoniadis:2018ywb,Rasanen:2018ihz,Antoniadis:2018yfq,Takahashi:2018brt,Tenkanen:2019jiq}, is the Palatini formalism \cite{Sotiriou:2008rp,DeFelice:2010aj,Borunda:2008kf,Olmo:2011uz,Kozak:2018vlp}, in which the metric tensor $g_{\mu \nu}$ and the connection $\Gamma_{\mu \nu}^{\lambda}$ are treated as independent variables.   Motivated by the multi-field scope of particle physics and the viability of Starobinsky inflation, in this paper we study the robustness of Starobinsky inflation in the presence of a scalar field non-minimally coupled to gravity both in the $\cal R$ and the ${\cal R}^2$ term. 

This paper is organized as follows. In Section 2 we present the general theoretical setup of the model at hand. In Section 3 we specialize our study and obtain numerical results for the observables for the case where the pre-existing scalar field is massless and its potential is zero. In Section 4 we do the same work with the addition of a mass term for the pre-existing scalar field. In Section 5 we identify $ \phi $ with the SM Higgs boson at the electroweak scale. We conclude and discuss potential extensions and future study of this model in Section 6.


\section{Theoretical setup} \label{Starobinsky-like two field inflation}

Our starting point is an inflationary model including a real scalar field $\phi$  non-minimally coupled to $F({\cal R})$ gravity. This model, is described by the action, in the $F({\cal R})$-frame,
\begin{equation}
\label{eqa}
 S=\int d^4x \sqrt{-\tilde{g}} \left[ F(\tilde{{\cal R}},\phi)  + \frac{1}{2} \tilde{g}^{\mu \nu} \partial_{\mu}\phi \partial_{\nu} \phi - U(\phi) \right] \, .
\end{equation}
For this theory we can obtain a better physical understanding by working in the Einstein frame. It can easily be seen that this theory is classically equivalent to 
\begin{equation}
\label{lagrang mult}
\begin{aligned}
S=&\int d^4x \sqrt{-\tilde{g}} \left[  F(\Phi,\phi)  + \psi (\Phi-\tilde{{\cal R}})+\frac{1}{2} \tilde{g}^{\mu \nu} \partial_{\mu}\phi \partial_{\nu} \phi - U(\phi) \right] \, ,
\end{aligned}
\end{equation}
as the equations of motion for the field $\psi$, which plays the role of a Lagrange multiplier, yield
\begin{equation}
\frac{\partial {\cal L}}{\partial \psi} = 0 \Rightarrow \Phi = \tilde{{\cal R}} \, .
\end{equation}
The corresponding equations of motion for $\Phi$ are then
\begin{equation}
 \frac{\partial {\cal L}}{\partial \Phi} = 0  \Rightarrow \frac{\partial F(\Phi,\phi)}{\partial \Phi} =-\psi \Rightarrow \Phi = \xi(\psi,\phi) \, .
 \label{xi}
\end{equation} 
Thus the action (\ref{lagrang mult}) can be written in the following form in the Jordan frame
\begin{equation}
\label{Jordan}
\begin{aligned}
S=&\int d^4x \sqrt{-\tilde{g}} \left[ -\psi \tilde{{\cal R}}+ F(\xi(\psi,\phi),\phi)  + \psi \xi(\psi,\phi) + \frac{1}{2} \tilde{g}^{\mu \nu} \partial_{\mu}\phi \partial_{\nu} \phi - U(\phi) \right] \, .
\end{aligned}
\end{equation}

In order to pass to the Einstein frame we perform a conformal transformation of the metric tensor $\tilde{g}_{\mu\nu}=g_{\mu\nu}/2\psi$ \cite{Birrell,Birrell2}. Under this transformation the Ricci scalar transforms as 
\begin{equation}
\label{Ricci}
{\cal \tilde{R}}=2\psi {\cal R}+\frac{3}{\psi} g^{\mu\nu}\partial_{\mu}\psi\partial_{\nu}\psi+6\psi^2 \nabla_{\mu}\left( \frac{\partial^{\mu}\psi}{\psi^2} \right) \, .
\end{equation} 
The action, after eliminating a total derivative, is given by
\begin{equation}
\label{eqav}
\begin{aligned}
S=&\int d^4x \sqrt{-g} \left[ -\frac{{\cal R}}{2} + \frac{3}{4} \left( \frac{\partial \psi}{\psi}\right)^2+ \frac{g^{\mu \nu}}{4\psi} \partial_{\mu}\phi \partial_{\nu} \phi - \frac{U(\phi)-\psi \xi(\psi,\phi) - F(\xi(\psi,\phi),\phi)}{4\psi^2} \right] \, .
\end{aligned}
\end{equation}
Finally, using the field redefinition $2\psi = e^{\sqrt{\frac{2}{3}}\rho}$, which leads to a canonical kinetic term for the field $\rho$, we obtain the following form for the action in the Einstein frame:
\begin{equation}
\label{eqg}
\begin{aligned}
S=&\int d^4x \sqrt{-g} \left[ -\frac{{\cal R}}{2} + \frac{1}{2} g^{\mu \nu} \partial_{\mu}\rho \partial_{\nu} \rho + \frac{1}{2} e^{-\sqrt{\frac{2}{3}}\rho} g^{\mu \nu} \partial_{\mu}\phi \partial_{\nu} \phi - V(\phi,\rho) \right] \, .
\end{aligned}
\end{equation}
In this the potential $V(\phi,\rho)$ is given by
\begin{equation}
\begin{aligned}
V(\phi,\rho)=& e^{-2\sqrt{\frac{2}{3}}\rho} \left( -\frac{1}{2}e^{\sqrt{\frac{2}{3}}\rho} \xi(\rho,\phi) - F(\xi(\rho,\phi),\phi)+U(\phi)  \right) \, .
\label{V general}
\end{aligned}
\end{equation}

An interesting model belonging to this class of theories is that proposed in \cite{Ketov} where
\begin{equation}\label{FRphi}
F({\cal R},\phi)= -\frac{1}{2} f(\phi) {\cal R} + \frac{1}{12M^2(\phi)}{\cal R}^2
\end{equation}
In this $f(\phi)$ and $M^2(\phi)$ are two generic functions, which should be positive defined in order to avoid ghosts. In this model one can easily find, using equation \eqref{V general}, that the potential $V(\phi,\rho)$ is given by the following expression
\begin{equation}
\label{pot1}
V(\phi,\rho)=e^{-2\sqrt{\frac{2}{3}}\rho} \left[ \frac{3}{4}M^2(\phi) \left( f(\phi)-e^{\sqrt{\frac{2}{3}}\rho} \right)^2 + U(\phi) \right] \, ,
\end{equation}
which however is in disagreement with the result derived in \cite{Ketov} \footnote{The potential found therein is of the form $V(\phi, \rho)=\frac{3}{4}M^2(\phi) \left[ f(\phi)-e^{\sqrt{\frac{2}{3}}\rho} \right]^2 +e^{2\sqrt{\frac{2}{3}}\rho}U(\phi)$.}. In the following we shall study the cosmological prediction of this model with the potential as given by  \eqref{pot1}.

The equations of motion for the two scalar fields $(\rho, \phi)$ in a spatially flat Friedman-Robertson-Walker spacetime
\begin{equation}
ds^2=dt^2-\alpha^2(t) d\vec{x}^2
\end{equation} 
assuming that the fields are only time-dependent, based on the observed homogeneity and isotropy of our Universe at large scales, following from the action \eqref{eqg} are given by
\begin{align}
\label{eoma}
\ddot{\rho}+3H\dot{\rho}+\frac{1}{\sqrt{6}}e^{-\sqrt{\frac{2}{3}}\rho}( \dot{\phi})^2 &=-V_{,\rho} \, ,
\\
\label{eomb}
\ddot{\phi}+3H\dot{\phi}-\sqrt{\frac{2}{3}}\dot{\rho}\dot{\phi} &=-e^{\sqrt{\frac{2}{3}}\rho} V_{,\phi}  \, ,
\end{align} 
where we denote $\dot{}\equiv d/dt$, $V_{,I}=\partial V/ \partial \phi^I$ with $I=\phi, \rho$ and $H=\dot{\alpha}/\alpha$ is the Hubble rate. The Einstein equations for the action \eqref{eqg} lead to the following Friedmann equations
\begin{align}
\label{frieda}
 3H^2 &=  \frac{\dot{\rho}^2}{2}+\frac{\dot{\phi}^2}{2}e^{-\sqrt{\frac{2}{3}}\rho}+V(\phi,\rho) \, ,
\\
\label{friedb}
 \dot{H} &=-\frac{1}{2} \left[ \dot{\rho}^2 +\dot{\phi}^2 e^{-\sqrt{\frac{2}{3}}\rho} \right] \, .
\end{align} 
In order to find the time-evolution of the fields $\rho$, $\phi$ and the scale factor we need just to solve the equations of motion \eqref{eoma}, \eqref{eomb} and \eqref{frieda}. The equation \eqref{friedb} is not independent but it is related to the other three equations of motion.

It is worth noting that the action \eqref{eqg} can be seen as a special case of the well-studied generalized non-linear sigma model of multifield inflation
\begin{equation}
\label{general}
S=\int d^4x \, \sqrt{-g} \left[ -\frac{R}{2}+\frac{1}{2}{\cal G}_{IJ}g^{\mu\nu}\partial_{\mu}\phi^I \partial_{\nu}\phi^J -V(\phi^I) \right] \, .
\end{equation}
In the above expression, the Latin indices account for the number of the fields of the theory and ${\cal G}_{IJ}$ is the metric tensor of the curved field space manifold. Then, in correspondence with the theory of Gravity, we have the definition for the covariant derivative:
\begin{equation}
\label{fieldcova}
{\cal D}_J A_I= A_{I,J}- \Gamma^{K}_{IJ}A_K \, ,
\end{equation}
with $A_I$ being a vector in the field space and $\Gamma^K_{IJ}$ being the corresponding Christoffel symbols in the curved field space, calculated by the expression
\begin{equation}
\label{Christof}
\Gamma^K_{IJ}=\frac{1}{2} {\cal G}^{KL} (\partial_I {\cal G}_{LJ}+\partial_J {\cal G}_{IL}-\partial_L{\cal G}_{IJ}) \, .
\end{equation}
The Einstein equations{} for the action \eqref{general} are of the usual form
\begin{equation}
\label{Einstein}
R_{\mu\nu}-\frac{R}{2}g_{\mu\nu}=T_{\mu\nu} \, ,
\end{equation}
where the stress tensor is given by the expression 
\begin{equation}
\label{Tmn}
T_{\mu\nu}=G_{IJ}\partial_{\mu}\phi^I\partial_{\nu}\phi^J-g_{\mu\nu}\left[\frac{G_{IJ}}{2}g^{\rho\lambda}\partial_{\rho}\phi^I\partial_{\lambda}\phi^J + V(\phi^I) \right] \, .
\end{equation}
The equations of motion for the fields $\phi^I$ are given by the expression
\begin{equation}
\label{fieldeq}
g^{\mu\nu}\phi^I_{;\mu;\nu}+g^{\mu\nu}\Gamma_{JK}^I \partial_{\mu}\phi^J \partial_{\nu}\phi^K-G^{IK}V_{,K}=0
\end{equation}
In this generalized model the slow-roll parameters are defined \cite{Kohri} as\footnote{In this work we assume slow-roll, slow-turn solutions. For highly curved field spaces the dominant behaviour can be of different nature (see e.g. \cite{Christodoulidis:2019mkj}).}
\begin{align}
\label{epsilon}
\epsilon &=\frac{1}{2}\frac{{\cal G}^{IJ} V_{,I}V_{,J}}{V^2} \, ,
\\
\label{eta}
\eta &=4\epsilon-\frac{{\cal D}_K V_{,J}{\cal G}^{KL}V_{,L}{\cal G}^{JM} V_{,M}}{\epsilon V^3} \, .
\end{align}
and should obey the ordinary (from one-field inflation models) slow-roll conditions
\begin{equation}
\label{slow}
\epsilon \ll 1 \, \, \, \, \, \, \, \, \, \, \text{and} \, \, \, \, \, \, \, \, \, \, \eta \ll 1 \, .
\end{equation} 
Also for this case of inflationary{} models, there have been derived some well known expressions{} for the spectral index and the tensor-to-scalar{} ratio that we measure today from the CMB{} spectrum. The definition of $n_s$ and $r$ happens in the context of cosmological perturbations. In the framework of this theory we have{} quantum fluctuations of the fields sourcing perturbations{} in the metric and vice versa. Let us{} briefly mention, using the description used in \cite{Sfakianakis,Sfakianakis2,Kaiser}, the basic concepts for the understanding of the quantities needed for the calculation of $n_s$ and $r$ in the concept of multi-field inflation. For a more complete{} view of this theory we recommend the reader to see these articles and the references therein \cite{Starobinsky:1986fxa,Mukha,Starobinsky:2001xq,Wands2007,Sfakianakis,Sfakianakis2,Kaiser,Riotto,Peterson,Peterson2,Gordon:2000hv,Basset,Malik,Lalak:2007vi,Langlois:2008mn,White:2012ya}. 

The results that we will  quote in the following are obtained by keeping only first order terms in the perturbative expansion of the spacetime-dependent fields, $\varphi^I(x^{\mu})$, around the time-dependent background fields, $\phi^I(t)$
\begin{equation}
\label{phipert}
\phi^I (x^{\mu})=\varphi^I (t)+\delta\phi^I (x^{\mu}) \, ,
\end{equation}
and of the spacetime metric around the spatially flat FRW background
\begin{equation}
ds^2=(1+2A) dt^2 - 2 \alpha (t) \partial_i B dx^i dt - \alpha^2 (t) [(1-2\psi ) \delta_{ij} + 2 \partial_i \partial_j E]dx^i dx^j \, .
\end{equation}
In order to have a gauge-invariant{} formalism and due to the fact that the fluctuations $\delta\phi^I $ do not transform covariantly{} under spacetime gauge transformations, it is necessary to introduce{} the gauge-invariant Mukhanov-Sasaki variables 
\begin{equation}
\label{Mukhanov-Sasaki}
Q^I=\delta\phi^I +\frac{\dot{\varphi}^I}{H}\psi \, .
\end{equation}
Using these variables we can rewrite the equations of motion \eqref{fieldeq} separating to background and first-order expressions as
\begin{equation}
\label{varphieom}
{\cal D}_t \dot{\varphi}^I+3H\dot{\varphi}^I+G^{IK}V_{,K}=0 \, ,
\end{equation}
and
\begin{equation}
\label{sasakieom}
{\cal D}^2_t Q^I+3H{\cal D}_t Q^I+\left[ \frac{k^2}{\alpha^2}\delta_J^I+ {\cal M}_J^I-\frac{{\cal D}_t}{\alpha^3}  \left( \frac{\alpha^3}{H}\dot{\varphi}^I\dot{\varphi}_I \right) \right] Q^J=0 \, ,
\end{equation}
where $k$ is the comoving wave number and the derivative ${\cal D}_t$ is defined as
\begin{equation}
{\cal D}_t A^I \equiv \dot{A}^I+\Gamma_{JK}^I A^J \dot{\varphi}^K \, ,
\end{equation}
while the mass-squared matrix ${\cal M}_J^I$ is defined as
\begin{equation}
\label{MIJ}
{\cal M}_J^I \equiv {\cal G}^{IK}({\cal D}_J {\cal D}_K V)-{\cal R}^I_{LMJ}\dot{\varphi}^L\dot{\varphi}^M 
\end{equation}
A useful quantity for the simplification of the analysis of the multi-field cosmological perturbations is the length of the velocity vector for the background fields given by
\begin{equation}
\label{sigdot}
\dot{\sigma}^2={\cal G}_{IJ}\dot{\varphi}^I \dot{\varphi}^J \, ,
\end{equation} 
from which one can define the adiabatic (curvature) and entropy (isocurvature) directions in the curved field space via the unit vectors
\begin{equation}
\label{unitvectors}
\hat{\sigma}^I \equiv \frac{\dot{\varphi}^I}{\dot{\sigma}} \, \, \, \, \, \, \, \, \, \, \text{and} \, \, \, \, \, \, \, \, \, \, \hat{s}^I\equiv \frac{\omega^I}{\omega} \, ,
\end{equation} 
where the turn-rate vector is given by $\omega^I(t) \equiv {\cal D}_t \hat{\sigma}^I$ and $\omega=|\omega^I| $. In a two-field model any vector in field space can be decomposed into components along these two unit vectors
\begin{equation}
\label{Aunit}
A^I=\hat{\sigma}^I\hat{\sigma}_J A^J+\hat{s}^I\hat{s}_J A^J
\end{equation}
Using this we can decompose the Mukhanov-Sasaki variables along adiabatic and entropy directions
\begin{equation}
Q_{\sigma}\equiv \hat{\sigma}_I Q^I \, \, \, \, \, \, \text{and} \, \, \, \, \, \, Q_s\equiv \hat{s}_I Q^I
\end{equation}
Using these new definitions of the Mukhanov-Sasaki variables and making{} a Fourier transformation of the form $\partial_i\partial^i F(t,x^i)=-(k^2/\alpha^2)F_k(t)$ we obtain from equation \eqref{sasakieom} the following two equations
\begin{equation}
\label{Qsigmaequation}
\ddot{Q}_{\sigma}+3H\dot{Q}_{\sigma}+\left[ \frac{k^2}{\alpha^2}+{\cal M}_{\sigma \sigma}-\omega^2-\frac{1}{\alpha^3}\frac{d}{dt}\left(\frac{\alpha^3 \dot{\sigma}^2}{H} \right) \right] Q_{\sigma}=2\left[ \frac{d}{dt}-\frac{\hat{\sigma}^I V_{,I}}{\dot{\sigma}}-\frac{\dot{H}}{H} \right] (\omega Q_s) \,
\end{equation}
and
\begin{equation}
\label{Qsequation}
\ddot{Q}_s+3H\dot{Q}_s+\left[ \frac{k^2}{\alpha^2}+{\cal M}_{ss}+3\omega^2 \right]Q_s=\frac{4\omega k^2}{\dot{\sigma} \alpha^2} \Psi \, ,
\end{equation}
with $\Psi$ the gauge-invariant Bardeen potential
\begin{equation}
\label{Bardeen}
\Psi \equiv \psi+\alpha^2 H \left( \dot{E}-\frac{B}{\alpha} \right) \, .
\end{equation}
Above we used the following definitions for the mass-squared matrix
\begin{equation}
\label{MssMsisi}
{\cal M}_{\sigma \sigma}\equiv \hat{\sigma}_I \hat{\sigma}^J {\cal M}_J^I \, \, \, \, \, \, \text{and} \, \, \, \, \, \, {\cal M}_{ss}\equiv \hat{s}_I \hat{s}^J {\cal M}_J^I \, .
\end{equation}
The effective mass of the isocurvature perturbations is defined by 
\begin{equation}
\label{isomass}
\mu_s^2 \equiv {\cal M}_{s s} +3 \omega^2\,.
\end{equation}
If  $ \mu_s^2/H^2 \gg  1 $, the isocurvature perturbations are heavy during slow-roll, so the growth of $Q_s$ is strongly suppressed. On the other hand  a tachyonic mass, $ \mu_s^2<0 \,,$ leads to the rapid amplification of the isocurvature modes.\\
The gauge-invariant curvature perturbation and the{} renormalized entropy perturbation are expressed in terms of the $Q_s$ and $Q_{\sigma}$ via the relations
\begin{equation}
\label{RcS} 
R\equiv \frac{H}{\dot{\sigma}}Q_{\sigma} \, \, \, \, \, \, \text{and} \, \, \, \, \, \, S \equiv \frac{H}{\dot{\sigma}}Q_s \, .
\end{equation}
In the long-wavelength and slow-roll limits these two functions can be found (from the differential equations of the Mukhanov-Sasaki variables) to obey the differential equations  
\begin{align}
\label{difR}
\dot{R}&=2\omega S + \mathcal{O}\left( \frac{k^2}{\alpha^2 H^2} \right) \\
\label{difS}
\dot{S}&=\beta (t) H S + \mathcal{O}\left( \frac{k^2}{\alpha^2 H^2} \right)
\end{align}
with the function $\beta (t)$ given by
\begin{equation}
\label{beta}
\beta (t)=-2 \epsilon - n_{ss}+n_{\sigma \sigma}-\frac{4 \omega^2}{3H^2} \, .
\end{equation}
In the above equation $n_{ss}$ and $n_{\sigma\sigma}$ are the slow-roll parameters defined by
\begin{equation}
\label{nssnsisi}
n_{ss}\equiv \frac{{\cal M}_{ss}}{V(\phi^I)} \, \, \, \, \, \, \text{and} \, \, \, \, \, \, n_{\sigma\sigma}\equiv 
\frac{{\cal M}_{\sigma\sigma}}{V(\phi^I)} \, .
\end{equation}
The solutions of the equations \eqref{difR} and \eqref{difS} can be written in the form
\begin{align}
\label{Rt}
&R(t)=R(t_*)+ T_{RS}(t_*,t)S(t_*) \, , \\ 
\label{St}
&S(t)=T_{SS}(t_*,t) S(t_*) \, ,
\end{align}
where
\begin{align}
\label{transfer2}
&T_{SS}(t_*,t)=\exp\left[{\int_{t_*}^{t} dt' \beta(t') H(t')}\right] \, ,
\\
\label{transfer1}
&T_{RS}(t_*,t)=\int_{t_*}^{t} dt' 2\omega(t') T_{SS}(t_*,t') \, ,
\end{align}
are the transfer functions which{} relate the gauge-invariant perturbations at the time $t_*$, when the perturbations of pivot scale during the inflationary era crossed outside the Hubble volume for the first time, to their{} value at some later time $t$.

The dimensionless primordial spectra ${\cal P}_R(k)$, ${\cal P}_S(k)$ and $C_{RS}(k)$ are defined at some time $t$ after inflation during the radiation-dominated era when all cosmological scales are still outside the horizon
\begin{align}
\label{CSR}
{\cal C}_{RS_*}(k,t_*) &\equiv \frac{k^3}{2\pi^2} \langle R(t_*) S^*(t_*) \rangle \, , \\
\label{PS}
{\cal P}_S(k,t) &\equiv \frac{k^3}{2\pi^2} \langle S(t) S^*(t) \rangle \,, \\
\label{PRa}
{\cal P}_R(k,t) &\equiv \frac{k^3}{2\pi^2} \langle R(t) R^*(t)  \rangle ={\cal P}_R(k,t_*)+2 T_{RS} {\cal C}_{RS_*}(k,t_*)+T^2_{RS} {\cal P}_S(k,t_*) \, .
\end{align}

Due to the fact that we are interested in the calculation of the power spectrum ${\cal P}_R(k)$ to first-order in the slow-roll parameters we keep only term to zeroth-order in ${\cal P}_R(k,t_*)$, ${\cal C}_{RS_*}(k,t_*)$ and ${\cal P}_S(k,t_*)$, where we have 
\begin{align}
{\cal C}_{RS_*}(k,t_*)&\simeq 0 \, , \\ 
{\cal P}_R(k,t_*)&={\cal P}_S(k,t_*)\simeq \left.\frac{V}{24\pi^2\epsilon}\right|_{t=t_*} \, .
\end{align}
Thus replacing back to \eqref{PRa} we obtain the equation which relates the power spectrum of curvature perturbations at time $t_*$ to its value a latter time $t$
\begin{equation}
\label{PRt}
{\cal P}_R(k,t)={\cal P}(k,t_*) \left[1+T^2_{RS}(t_*,t) \right]
\end{equation}
The spectral index can be determined via the power spectrum of the{} curvature perturbations from the expression
\begin{equation}
\label{specta}
n_{s}(t)=1+ \frac{\partial \ln {\cal P_R}}{\partial \ln k}\simeq 1+\frac{d\ln V}{d\ln k}-\frac{d \ln \epsilon}{d \ln k}+\frac{d \ln(1+T^2_{RS})}{d \ln k} \, .
\end{equation}
Using the fact that the derivative with respect to $\ln k$ can be converted to a time derivative via the relation $d \ln k/dt=(1-\epsilon)H$ and that for the second and the third term of \eqref{specta} holds true
\begin{equation}
\frac{d \ln V}{d \ln k}\simeq -2 \epsilon \, \, \, \, \, \, \text{and} \, \, \, \, \, \, \frac{d \ln \epsilon}{d \ln k}=4\epsilon-2n_{\sigma\sigma} \, ,
\end{equation}
we find  that  \eqref{specta} takes the form
\begin{equation}
\label{spect}
n_{s}(t)=n_s(t_*)-\left(\frac{2\omega(t_*)}{H(t_*)}+\beta(t_*) {\cal T}_{RS}(t_*,t) \right) \sin(2\Delta)
\end{equation}
where we have used the definitions
\begin{align}
\label{Delta}
&\cos (\Delta) \equiv \frac{T_{RS}}{(1+T_{RS}^2)^{1/2}} \, , \\
\label{spectral}
&n_s(t_*)\equiv 1-6\epsilon(t_*)+2 \eta_{\sigma\sigma}(t_*) \, .
\end{align}

As it regards the tensor-to-scalar ratio, $r$, it can be determined{} from the ratio between the power spectrum of the tensor perturbations and that of the curvature perturbations, meaning
\begin{equation}
\label{ratio}
r\equiv \frac{P_T(k,t)}{P_R(k,t)} \, .
\end{equation} 
Due to the vanish of $T^i_j$ terms for $i \neq j$ to first-order the tensor perturbations evolve in the same way as in the single-field models and thus ${\cal P}_T$ is given by the expression
\begin{equation}
\label{PT}
{\cal P}_T(k,t_*) \equiv \frac{k^3}{2\pi^2} \langle \delta g_{\mu\nu} \delta g^{\mu\nu *} \rangle \simeq 8 \left.\left(\frac{H}{2\pi}\right)^2 \right|_{k=\alpha H}=\left.\frac{2V}{3\pi^2}\right|_{t=t_*}
\end{equation}
Thus for the tensor-to-scalar ratio in two-field models from equations \eqref{PRt} and \eqref{PT} we obtain
\begin{equation}
\label{rnew}
r \simeq \frac{16 \epsilon(t_*)}{1+T^2_{RS}} \, .
\end{equation}
From the above expression it is{} obvious that if the maximum value $r_{\text{max}}=16\epsilon(t_*)$ is in agreement with the observed{} constraint \eqref{obs}, there is no need of calculation of the transfer functions, as far as  $r$ is concerned.

For the system at hand one can easily infer, by comparing \eqref{eqg} and \eqref{general}, identifying $ \phi^1 $ by $ \rho $ and $ \phi^2 $ by $ \phi $, that the metric tensor ${\cal G}_{IJ}$ has the following form
\begin{equation}
\label{GIJ}
{\cal G}_{IJ}=\left( \begin{matrix}
1 & 0 \\
0 & e^{-\sqrt{\frac{2}{3}}\rho}
\end{matrix} \right) \, .
\end{equation} 
Then, by the use of \eqref{Christof} we find that the only{} non-zero components of $\Gamma^K_{IJ}$ are
\begin{equation}
\label{Gamma}
\Gamma^{\phi}_{\phi\rho}=\Gamma^{\phi}_{\rho\phi}=-\sqrt{\frac{1}{6}} \, \, \, \, \, \, \, \, \, \, \text{and} \, \, \, \, \, \, \, \, \, \, \Gamma^{\rho}_{\phi\phi}=\sqrt{\frac{1}{6}}e^{-\sqrt{\frac{2}{3}}\rho} \, .
\end{equation} 
Thus, using the definitions \eqref{epsilon}, \eqref{eta} and \eqref{fieldcova} for our case, we are leaded to the following expressions for the slow-roll parameters
\begin{align}
\label{slowepsilon}
&\epsilon = \frac{\left(V_{,\rho}\right)^2}{2V^2}+ e^{\sqrt{\frac{2}{3}}\rho}\frac{\left(V_{,\phi}\right)^2}{2V^2}\, , \\
\label{sloweta}
&\eta =4\epsilon-\frac{1}{\epsilon V^3}\Big( V_{,\rho\rho}V^2_{,\rho}+2e^{\sqrt{\frac{2}{3}}\rho}V_{,\rho\phi}V_{,\phi}V_{,\rho}+\sqrt{\frac{1}{6}}e^{\sqrt{\frac{2}{3}}\rho}V_{,\phi}^2 V_{,\rho}+e^{2\sqrt{\frac{2}{3}}\rho}V_{,\phi\phi}V_{,\phi}^2 \Big) \, , \\
\label{sloweta2}
& \eta_{\sigma\sigma}=\Big(( \hat{\sigma}^{\rho})^2 \frac{\partial^2 V}{\partial \rho^2}+( \hat{\sigma}^{\phi} )^2 \frac{\partial^2 V}{\partial \phi^2}+2\hat{\sigma}^{\rho}\hat{\sigma}^{\phi} \frac{\partial^2 V}{\partial \phi \partial \rho}+\frac{2\hat{\sigma}^{\rho}\hat{\sigma}^{\phi}}{\sqrt{6}}\frac{\partial V}{\partial \phi}-\frac{(\hat{\sigma}^{\phi})^2}{\sqrt{6}} e^{-\sqrt{\frac{2}{3}}\rho}\frac{\partial V}{\partial \rho} \Big) \frac{1}{V} \, ,
\end{align}
where the unit vectors $\hat{\sigma}^{\phi}$, $\hat{\sigma}^{\rho}$ appearing in the equations above are
\begin{equation}
\label{sigmas}
\hat{\sigma}^{\rho}=\frac{\dot{\rho}}{\sqrt{\dot{\rho}^2+e^{-\sqrt{\frac{2}{3}}\rho}\dot{\phi}^2}} \quad \text{and} \quad \hat{\sigma}^{\phi}=\frac{\dot{\phi}}{\sqrt{\dot{\rho}^2+e^{-\sqrt{\frac{2}{3}}\rho}\dot{\phi}^2}} \, .
\end{equation} 
So far we have not specified the functions $ U(\phi) $, $ f(\phi) $ and $ M(\phi) $ defining the function $ F(\tilde{{\cal R}},\phi) $, \eqref{FRphi}, which defines the model given in \eqref{eqa}.  In order to proceed to predictions of the inflationary observables within the context of this type of models in the following sections, and for different forms of the potential $U(\phi)$, we choose the functions $M^2(\phi)$ and $f(\phi)$ to be the same as those employed in \cite{Ketov},
\begin{equation}
\label{M,F}
M^2(\phi)=M^2(1+\beta \phi^2) \quad \text{and} \quad f(\phi)=1+\alpha \phi^2 \, .
\end{equation} 
In these $\alpha$ and $\beta$ are some constants signaling deviations from Starobinsky model. We mainly focus on the parameter regime where $ \alpha, \beta >0. $ Negative values for the parameters $ \alpha $ and $ \beta $ are also acceptable provided that the functions $ M^2(\phi) $ and $ f(\phi) $ remain positive to avoid ghosts.

In the next sections we will be concentrated on the calculation of the spectral index $ n_s $ and the tensor-to-scalar ratio $ r $ in the pivot scale. The main difference{} of our two-field inflationary{} model against{} the Starobinsky model is the presence{} of isocurvature perturbations. Those perturbations turn out to be very small and{} undetectable so far. So, we did not examine primordial non-Gaussianities in our deformation of Starobinsky inflationary model, because we expect it to be negligible, like in the pure Starobinsky model of inflation{} and thus in agreement with the constraints \cite{Planck,Planck2}. Also, for all the cases which will be studied in the following sections we found that the effective mass, $ \mu_s$, of the isocurvature perturbations is small and negative and thus the formalism we presented in this section is applicable.


\section{The Case $U(\phi)=0$}

\begin{figure}[t]
\begin{center}
\includegraphics[width=8cm]{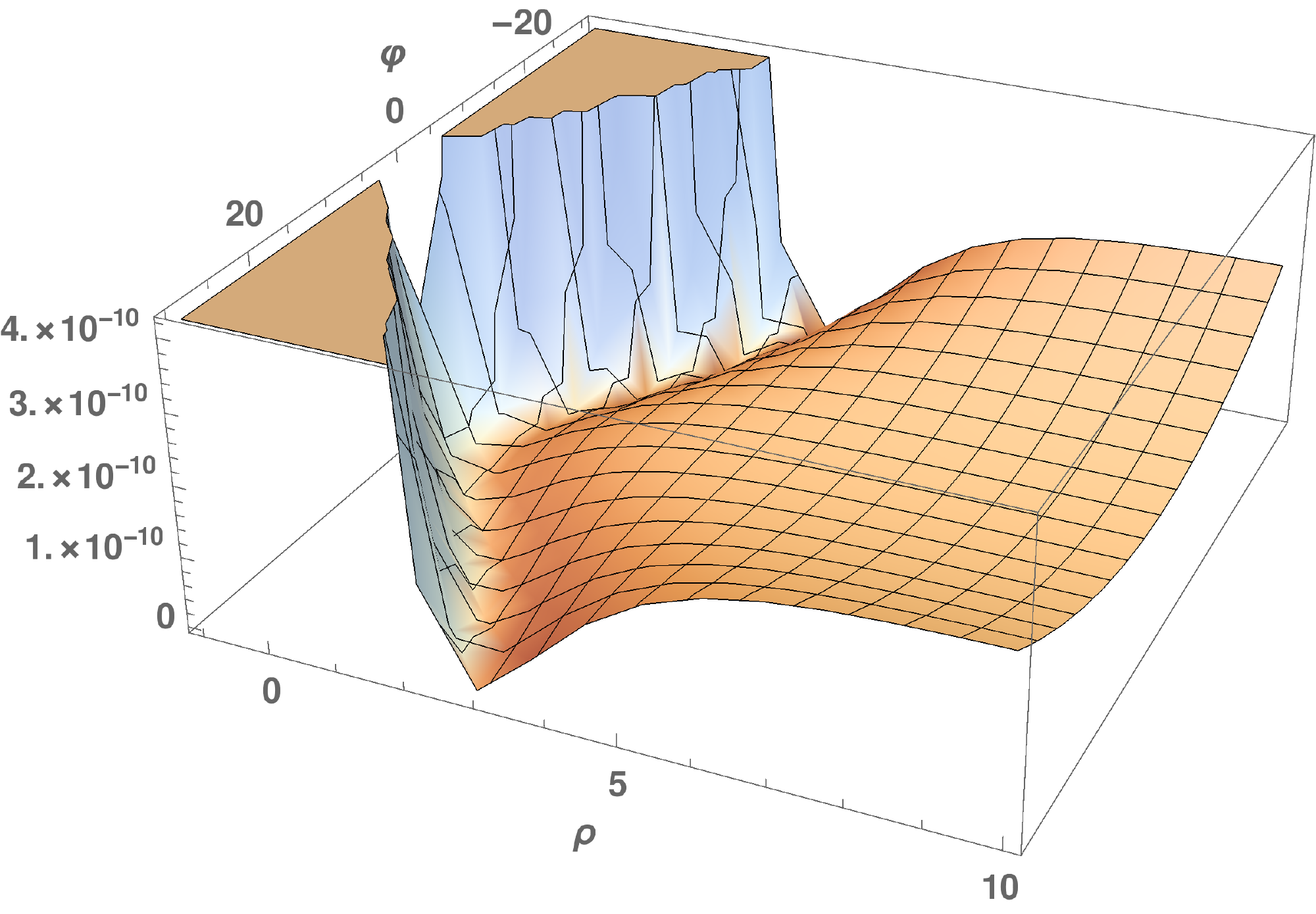}\caption{The potential $ V(\phi,\rho) $ for $ \alpha=0.01, $ $ \beta=0.001 $ and $ M=1.3\cdot 10^{-5}. $ }\label{fig1}
\end{center}
\end{figure}

In this section we review the special case, already studied in \cite{Ketov}, when $\phi$ is a free massless field, in the $F(\cal{R})$-frame. We have $U(\phi)=0$ and thus the scalar potential $V(\phi,\rho)$ becomes 
\begin{equation}
\label{pot0}
V(\phi,\rho) = \frac{3}{4} e^{-2\sqrt{\frac{2}{3}}\rho}\,M^2 (1+\beta \phi ^2) \left(1+\alpha \phi ^2 -e^{\sqrt{\frac{2}{3}}\rho}\right)^2 \, ,
\end{equation} 
which is semi-positive definite. From the above expression we observe that for $\alpha=\beta=0$ we retain the Starobinsky potential, as we expected. For non-zero values of $\alpha$ and $\beta$ we have a deformation of the potential in the $\phi$ direction, leading to the stabilization of the field $\phi$. In the case where only $ \alpha=0\,, $ the parameter $ \beta $ cannot grow well beyond 50. Actually, the marginal case where  $ \beta=50\,, $ is marginally out due to the non agreement with the $ n_s $ data. The increase of the parameter $ \beta $ does not have a significant effect on the value of $ r $. On the other hand, if $ \beta=0\,, $ then $ \alpha $ cannot be bigger than $ \sim 10^{-3}\,. $ Also, according to our numerical calculations if both $\alpha$ and $\beta$ grow well beyond 10 the inflation becomes again non-viable for the same reason, namely our model fails to get the observed value of $n_s$. From now on and in the following, for the parameters $ \alpha, \beta $ and $M$ we choose for definiteness the values $ \alpha=0.01, $ $ \beta=0.001 $ and $ M=1.3\cdot 10^{-5}$, like in the pure Starobinsky model. In Figure \ref{fig1} we show the profile of the potential for this choice of parameters. It is worth observing that in this potential we do not have a unique Minkowski vacuum but rather a valley of Minkowski vacua on the contour determined by the points $ (\phi_{\text{min}},\rho_{\text{min}})=(c,\sqrt{\frac{3}{2}}\ln(1+\alpha c^2)), $ where $ c\in {\rm I\!R}. $ As we shall see below, the choice of the initial values of the fields $ \phi $ and $ \rho $ lead us to the vacuum $ (\phi_{\text{min}},\rho_{\text{min}})=(0,0)$. The choice of different initial conditions will have as a result an altered vacuum which can lead us to successful inflation.

\begin{figure}[t]
\begin{center}
\includegraphics[width=7cm]{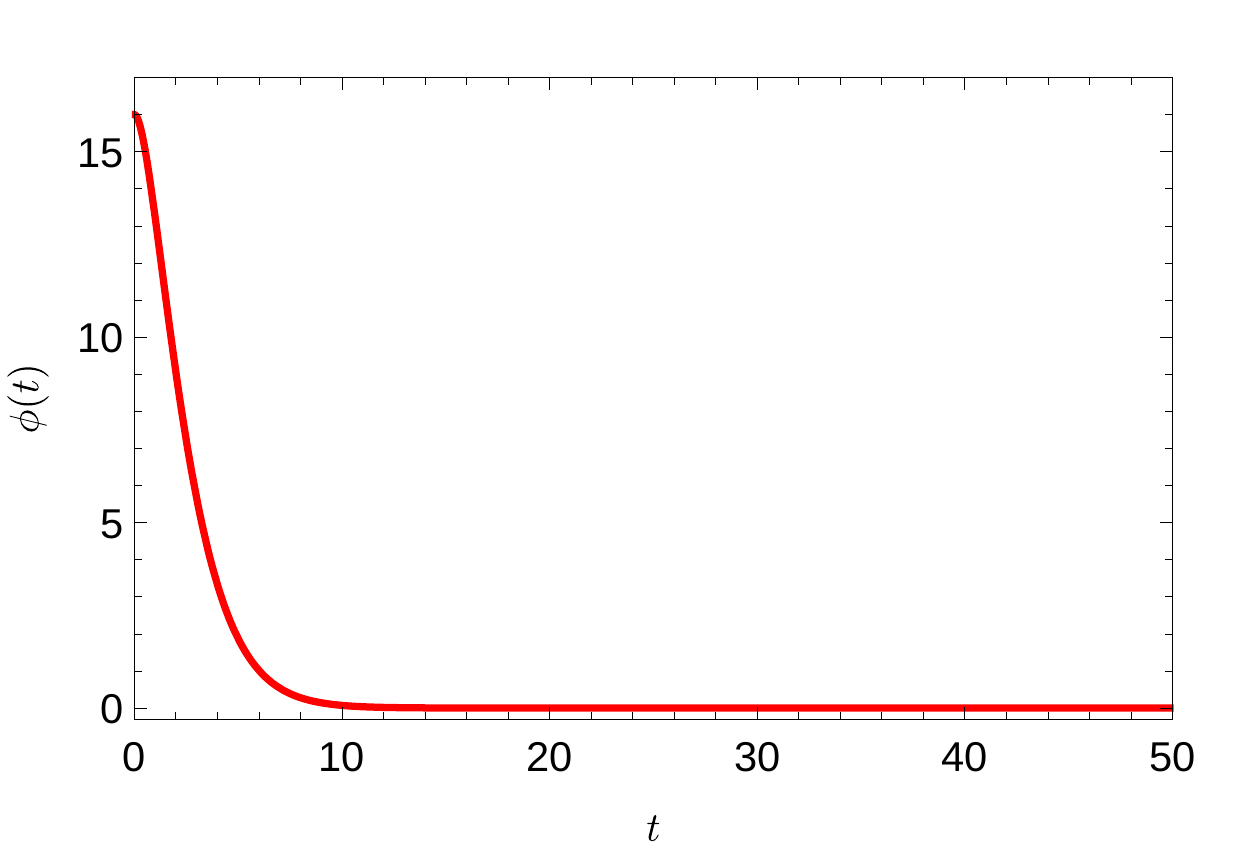}
\includegraphics[width=7cm]{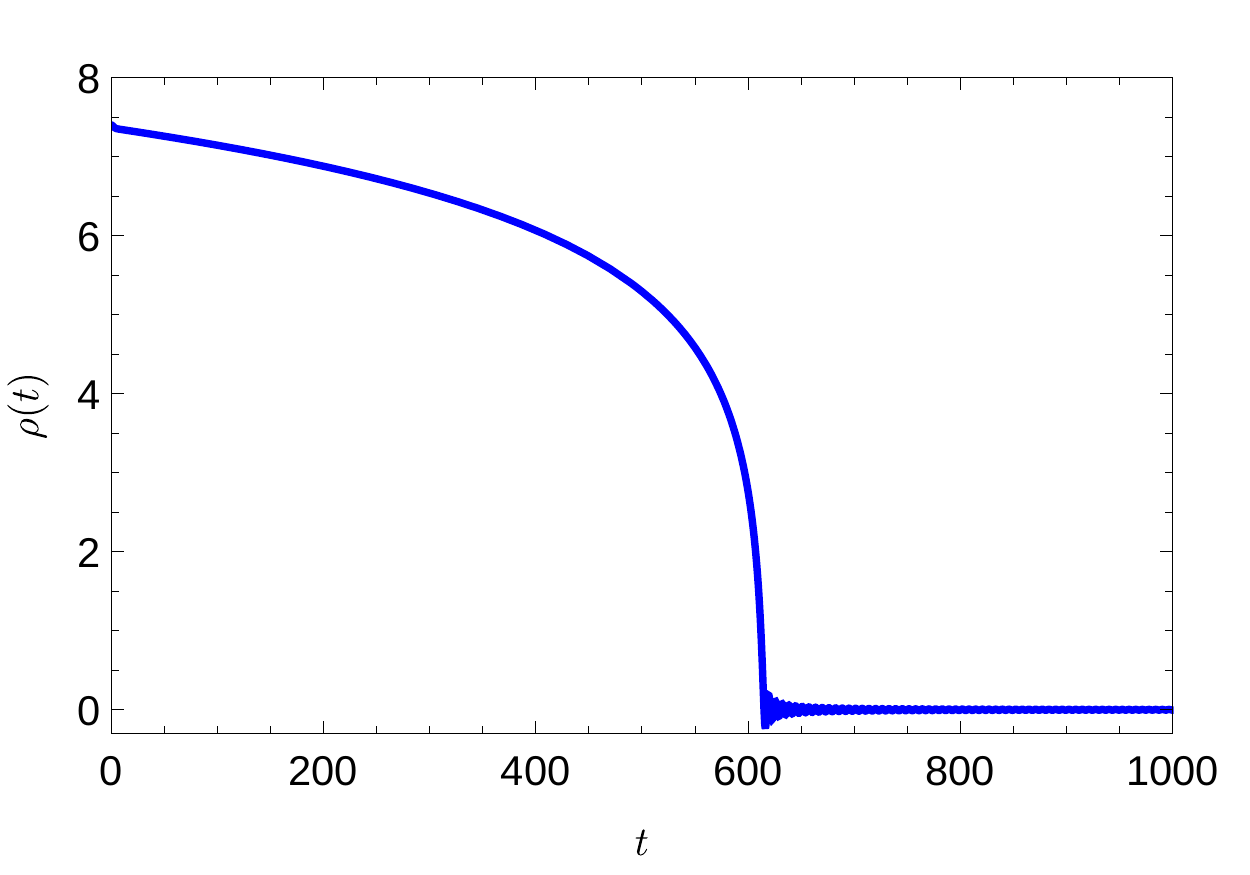}\caption{Left: The field $ \phi(t) $ as a function of time. Right: The field $ \rho(t) $ as a function of time.}\label{fig2}
\end{center}
\end{figure}

\begin{figure}[t]
\begin{center}
\includegraphics[width=7.5cm]{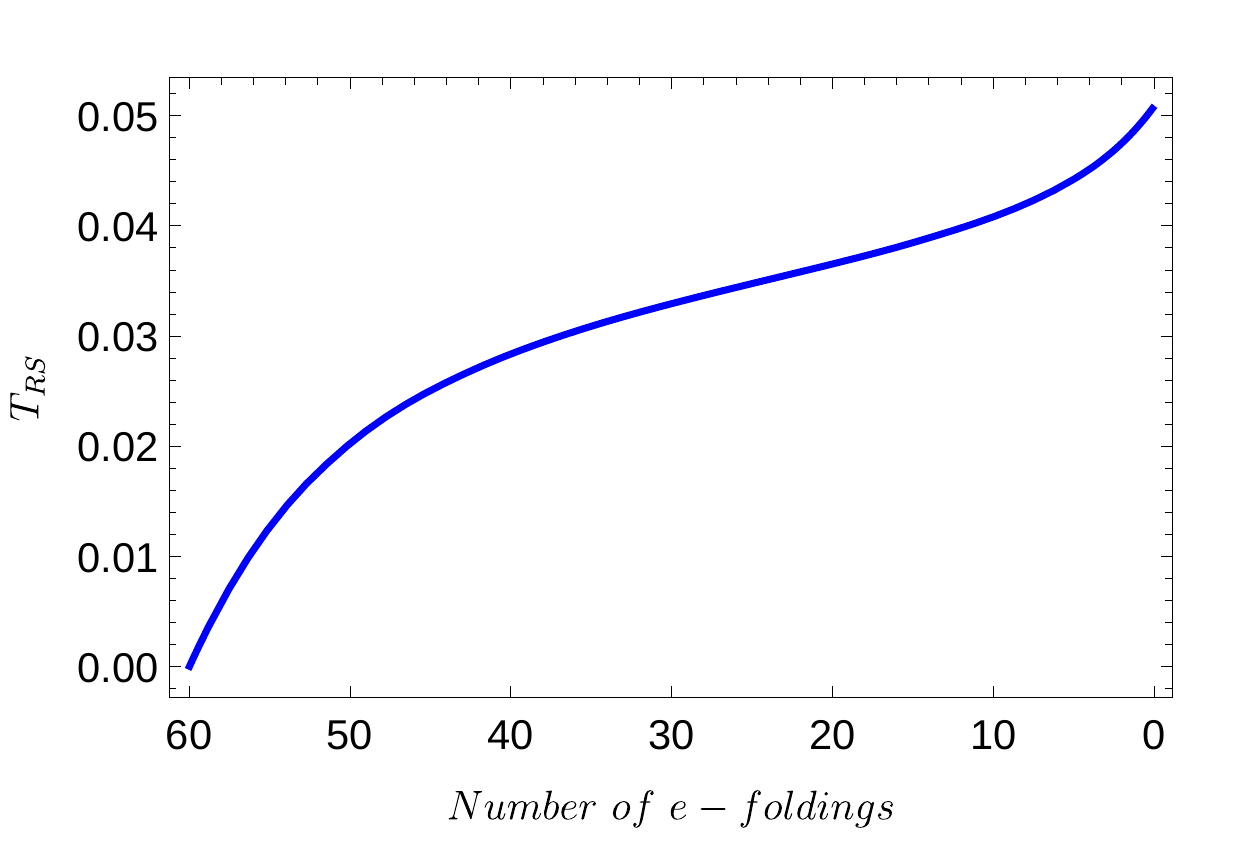}\caption{The transfer function $T_{RS}$ as a function of number of e-foldings.}\label{TRS}
\end{center}
\end{figure}

\begin{figure}[t]
\begin{center}
\includegraphics[width=7.2cm]{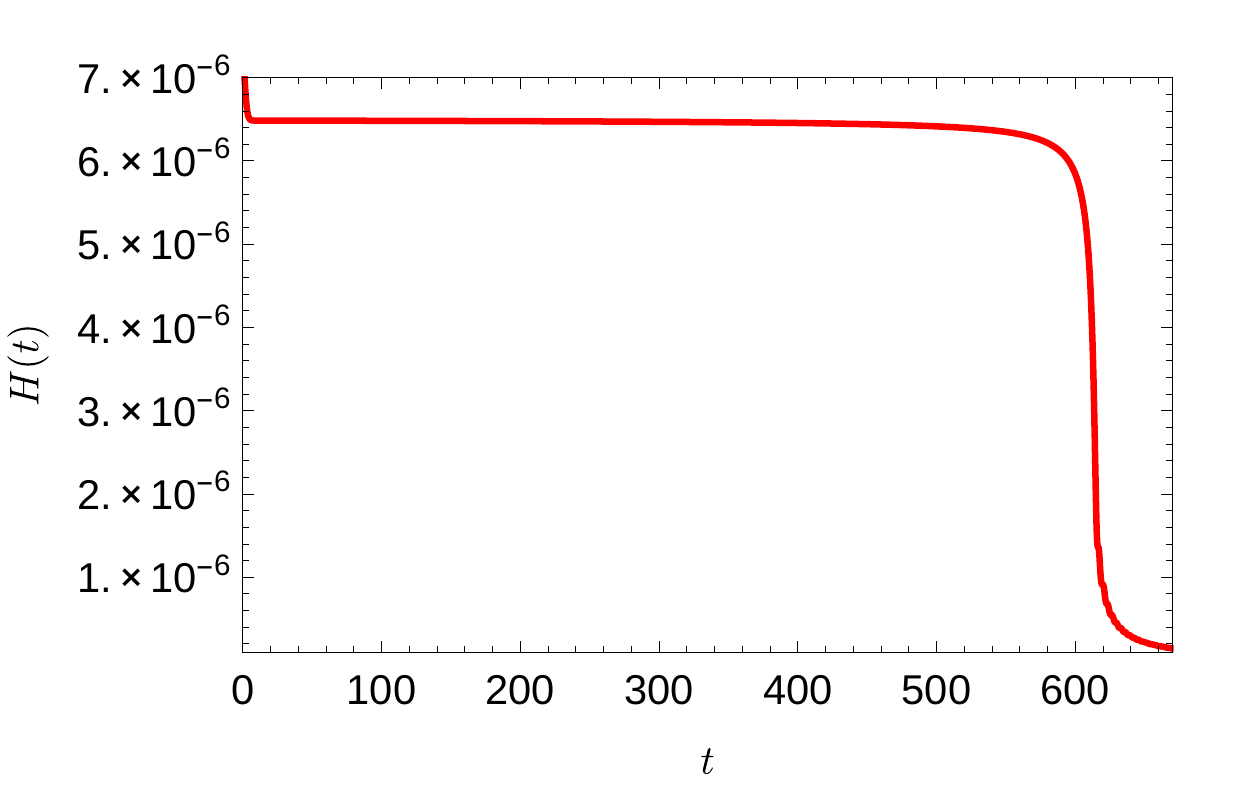}
\includegraphics[width=7cm]{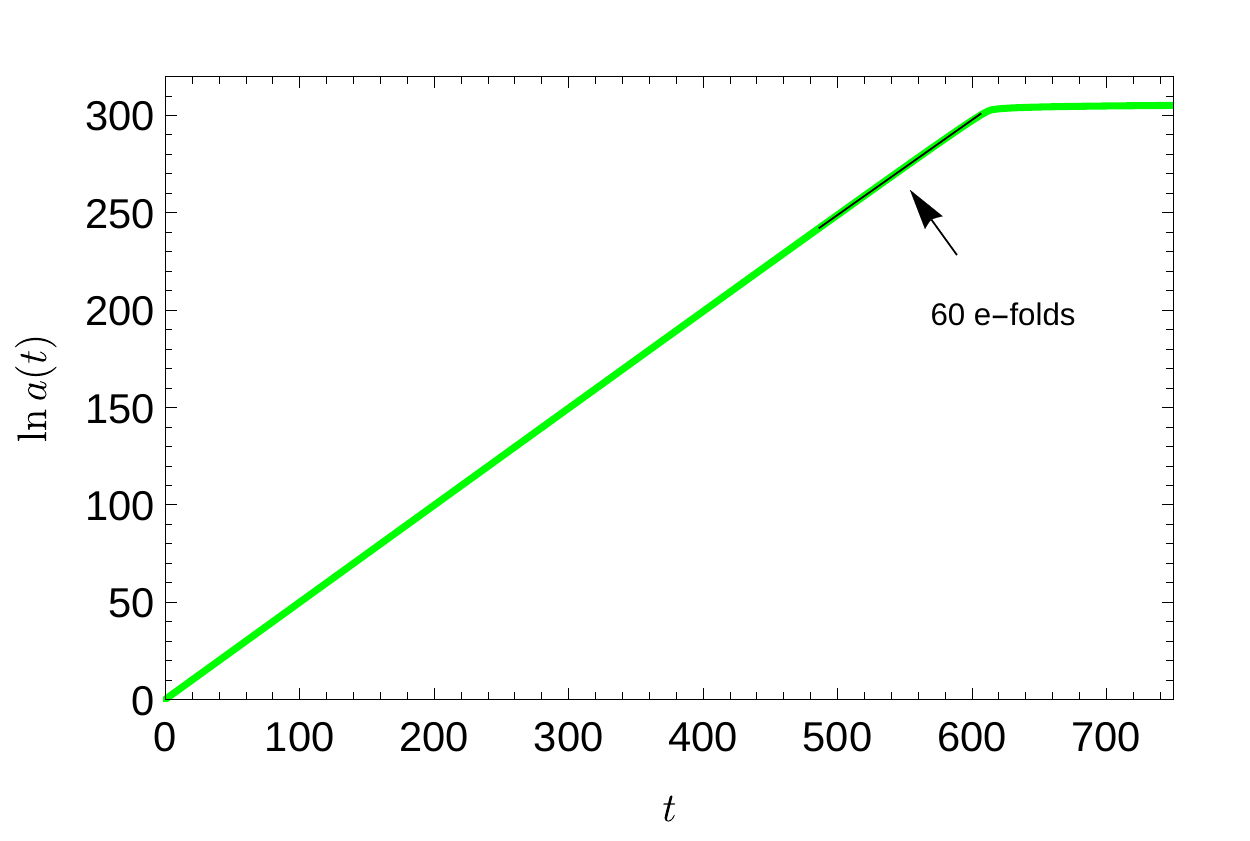}\caption{Left: The Hubble parameter as a function of time. Right: The logarithm of the scale factor $a(t)$. With the black line is denoted the time of the last 60 e-folds before the end of inflation. }\label{fig4}
\end{center}
\end{figure}

\begin{figure}[t]
\begin{center}
\includegraphics[width=7cm]{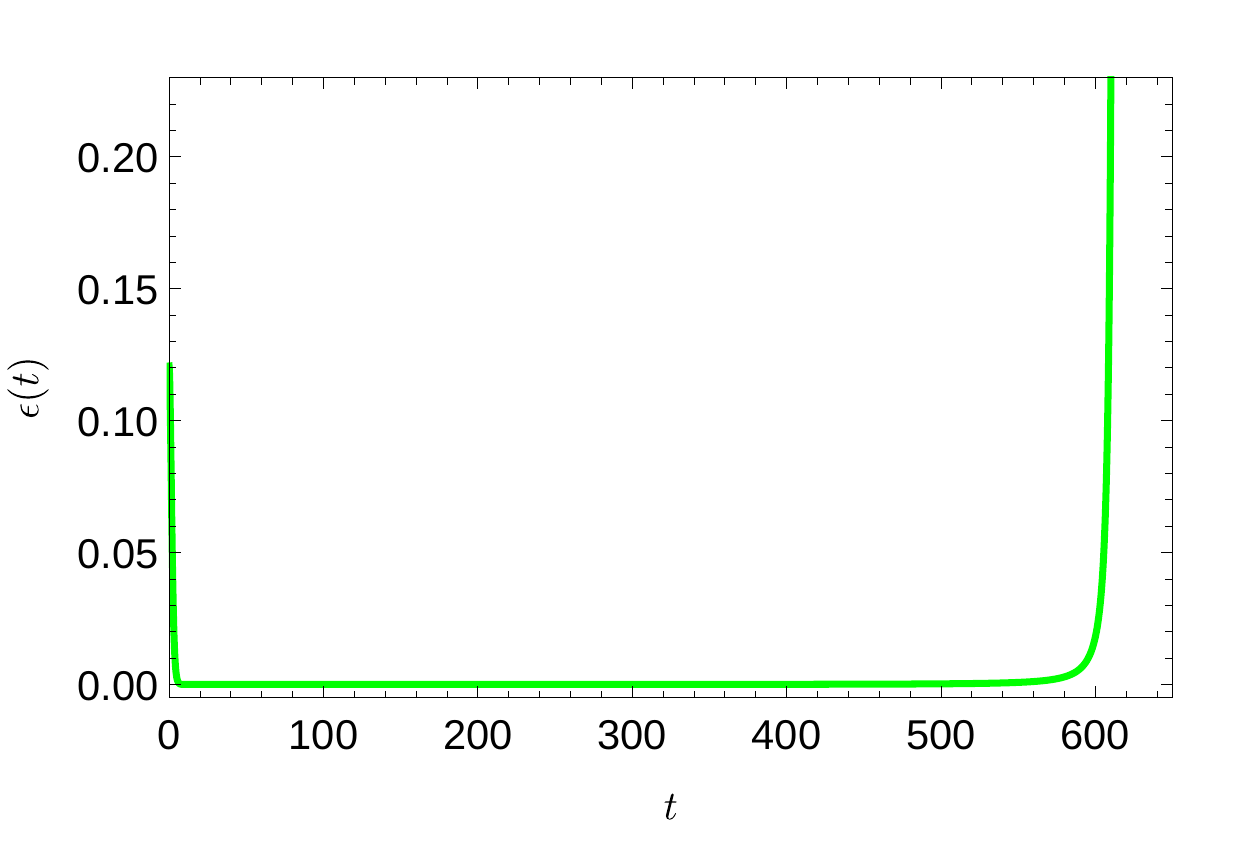}
\includegraphics[width=7cm]{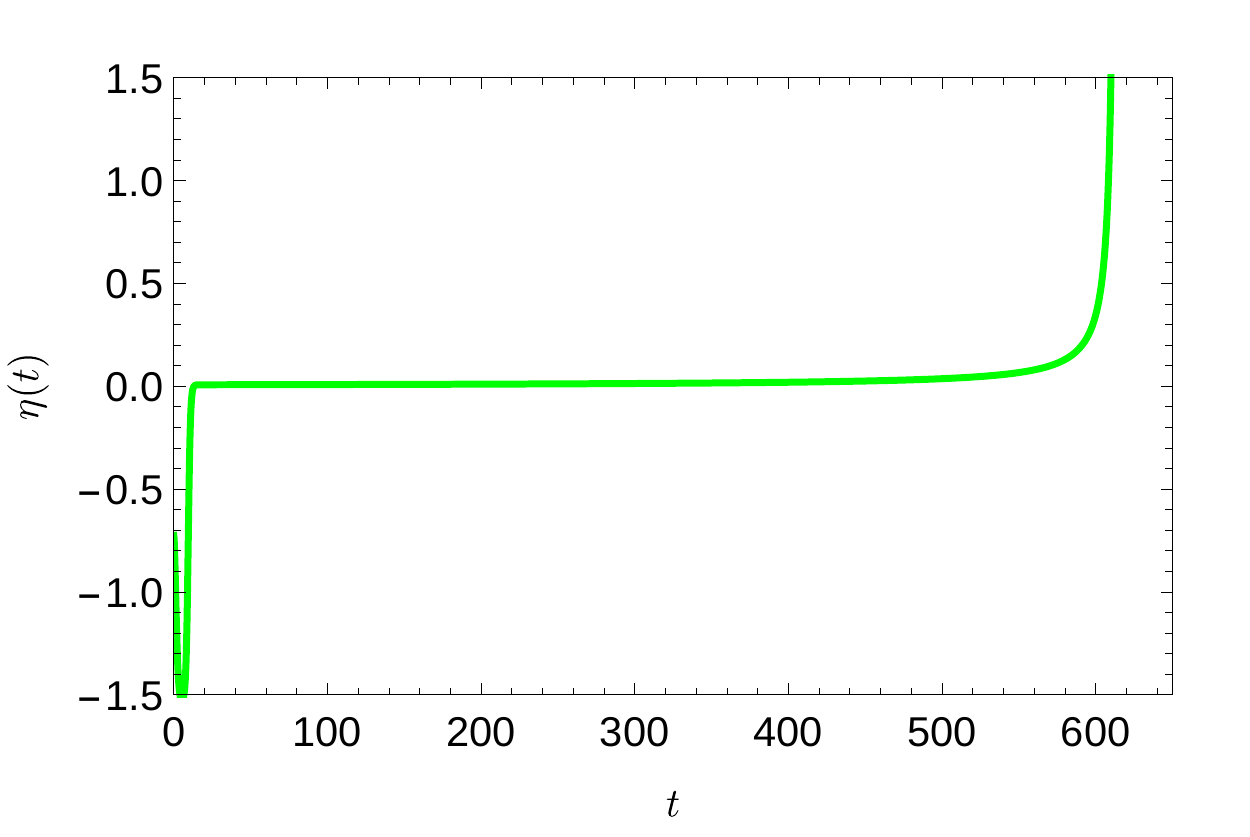}\caption{Left: The running of slow-roll parameter $ \epsilon $. Right: The running of slow-roll parameter $ \eta $.}\label{fig5}
\end{center}
\end{figure}

In order to calculate the scalar spectral index and the tensor-to-scalar ratio in the pivot scale, we solve numerically the equations of motion (\ref{eoma}) , (\ref{eomb}) and (\ref{frieda}) to find the time evolution of the fields $\phi$ and $\rho$. Our results are shown in Figure \ref{fig2}. In all graphs displayed the time $t$ is normalized by $M$. From Figure \ref{fig2} we can see that at the end of inflation the field $ \rho $ oscillates near in its approach to the minimum of the potential, whereas the field $ \phi $ does not. In fact $ \phi $ drops off rapidly. The oscillation of the field $ \rho $ can be seen after the time $t=605$. Using the numerical results for $\phi$ and $\rho$, we then solve the equation (\ref{frieda}), find the time dependence of the scale factor $a(t)$ and plot the time evolution of the Hubble parameter and the number of e-foldings, as shown in Figure \ref{fig4}.  From the bottom diagram of Figure \ref{fig4} we can find the pivot scale used for the calculation of the spectral index and the tensor-to-scalar ratio. Furthermore, using the relations (\ref{slowepsilon}) and (\ref{sloweta}) we calculate the slow-roll parameters $\epsilon$ and $\eta$, respectively, and we present their behaviour as a function of time in Figure \ref{fig5}. From that figure we can see that the slow-roll parameters $ \epsilon $ and $ \eta $ indeed obey the slow-roll conditions (\ref{slow}). Finally, using relations \eqref{spectral} and \eqref{rnew} we calculate the spectral index $ n_s $ and the max value of the tensor-to-scalar ratio in the pivot scale. For $ N\simeq 50-60 $ we find that\footnote{The calculation of $n_s$ and $ r $ has been made between 50 and 60 e-foldings, so the $ \pm  $ or $ \mp $ signs correspond to 50 (down sign) and 60 (upper sign) e-foldings respectively.}
\begin{equation}
\label{spectralandtensro}
n_s=0.965\pm 0.004 \quad \text{and} \quad r= 0.0037 \mp 0.0007 \, .
\end{equation} 
These values for $ n_s $ and $r$ are in agreement with the constraints (\ref{obs}) and thus lead to viable inflation. Scales of cosmological interest first crossed the Hubble radius between $50$ and $60$ e-foldings before the end of inflation. As indicated in Figure \ref{TRS}, $ T_{RS} $ remains small between $ N=50\,\, \text{and}\,\, 60, $ so corrections to $ n_s $ and $ r $ from $ T_{RS} $ remain negligible.

It is worth pointing out, that trying different values for the initial conditions of the fields we observed that for bigger values of $\phi$ we have smaller duration of the inflationary period (number of e-foldings) but always in agreement with the observational constraints \eqref{obs}. In fact we have a wide range of allowed initial values for the field $\phi$, for every initial value of $\rho$, to obtain at least $50-60$ e-foldings of inflation. Therefore we do not have fine tuning for the initial conditions since a wide range of them leads to realistic inflation.

\section{The Case $U(\phi)=\frac{1}{2}m^2 \phi^2$}

Now we make a simple but physical extension of the calculation we made in the previous section, by adding a mass term for the field $\phi$ in our action. The scalar potential $V(\phi,\rho)$ in this case takes the form
\begin{equation}
\label{4.1}
V(\phi,\rho)= e^{-2\sqrt{\frac{2}{3}}\rho}\left[\frac{3}{4} \,M^2 (1+\beta \phi ^2) \left(1+\alpha \phi ^2 -e^{\sqrt{\frac{2}{3}}\rho}\right)^2 + \frac{1}{2}m^2 \phi^2\right] \, .
\end{equation}

This model for the case of $\alpha=\beta=0$ was studied in references \cite{vandeBruck:2015xpa,Kohri} where it was found that it yields viable inflation without the need of fine-tunning. Specifically, in \cite{Kohri}, using the $\delta N$ formalism, the non-gaussianity  was calculated and was found to be in agreement with the observed constraints \cite{Planck,Planck2}. Thus our expectation of feasible non-gaussianity for the case $\alpha=0.01$ and $\beta=0.001$ is strengthened. In this model, values for  $\alpha$ and $\beta$ greater than 10 but less than $10^2$ are also acceptable as we have checked numerically.

\begin{figure}[t]
\begin{center}
\includegraphics[width=8cm]{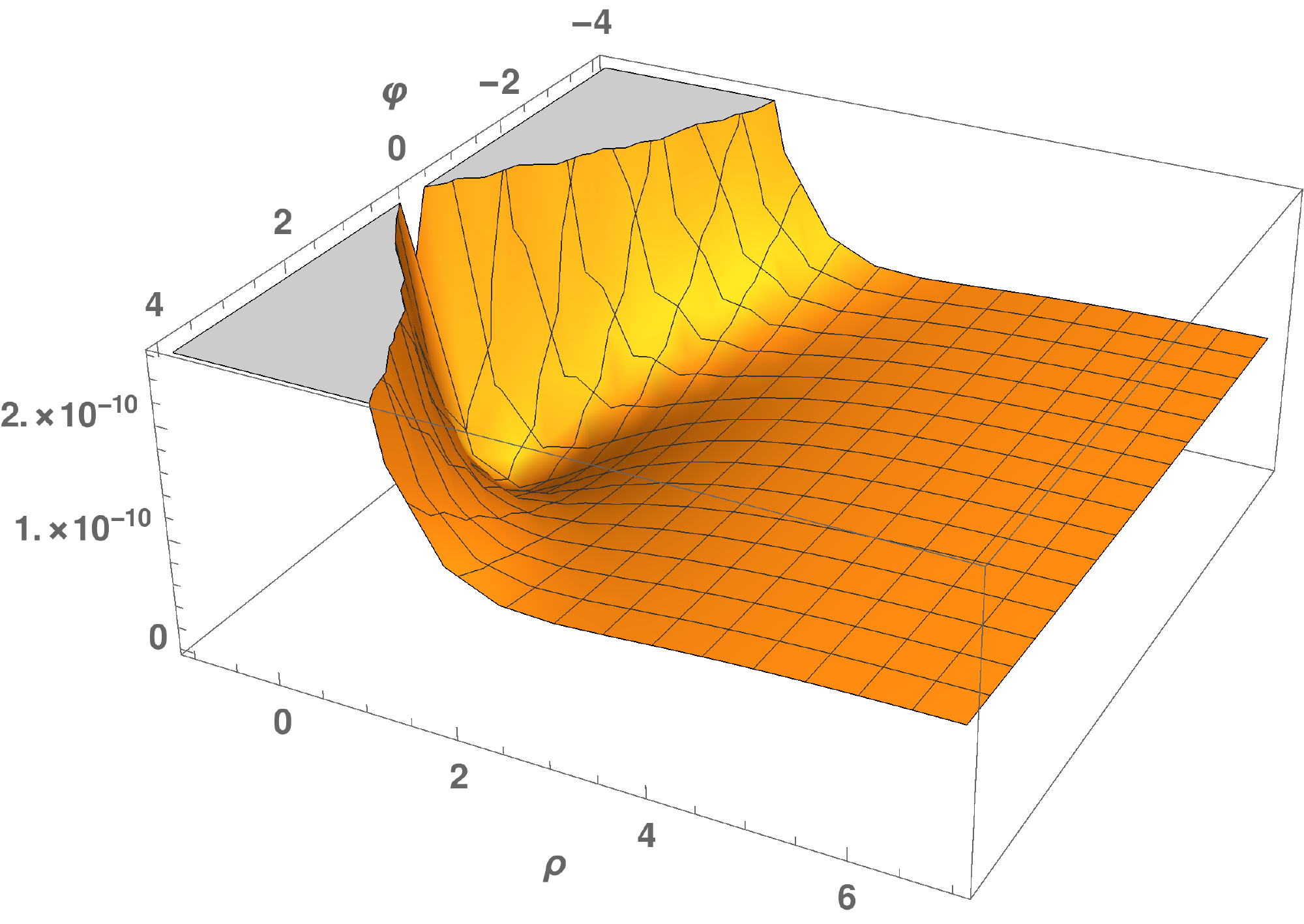}\caption{The potential $ V(\phi,\rho) $ for $\alpha=0.01$, $\beta=0.001$ and $m=M=1.3\cdot 10^{-5}$.}\label{pot}
\end{center}
\end{figure}

\begin{figure}[t]
\begin{center}
\includegraphics[width=7cm]{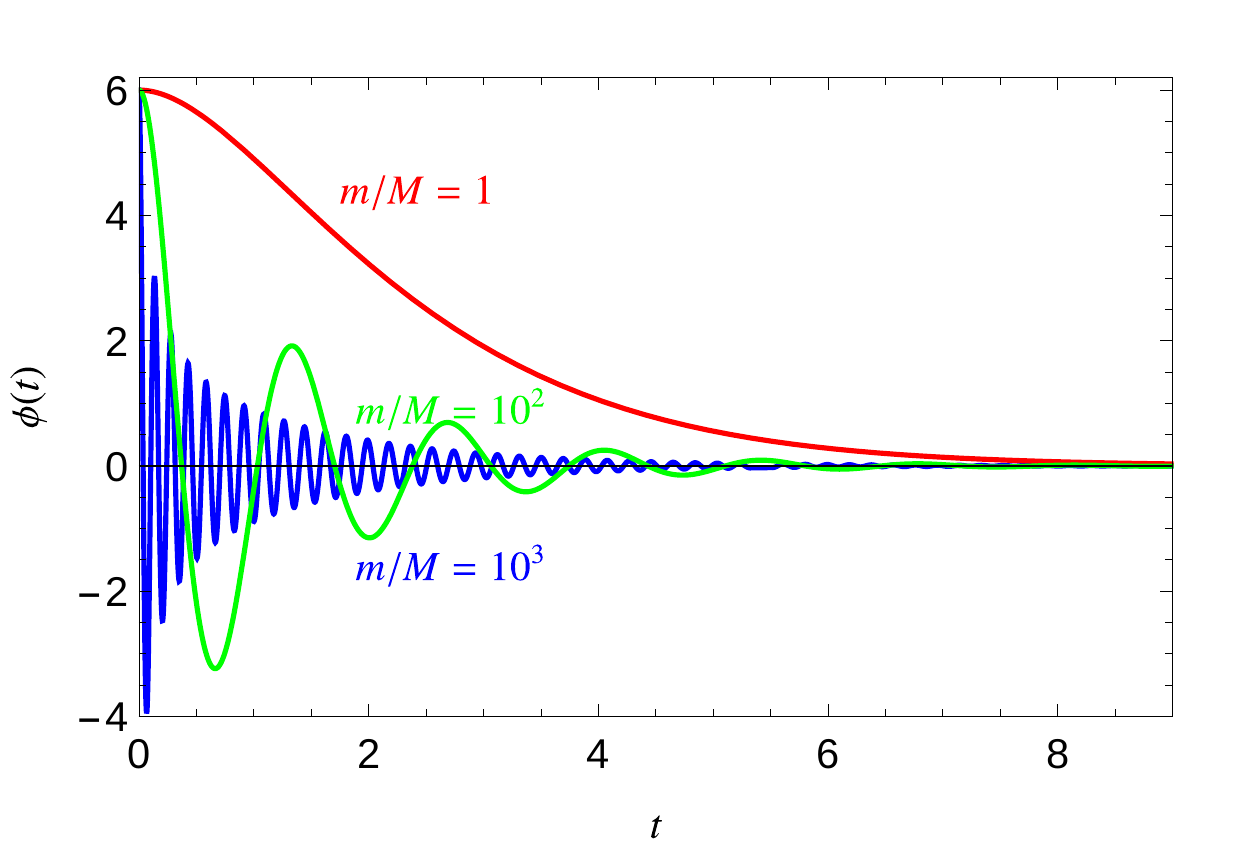}
\includegraphics[width=7cm]{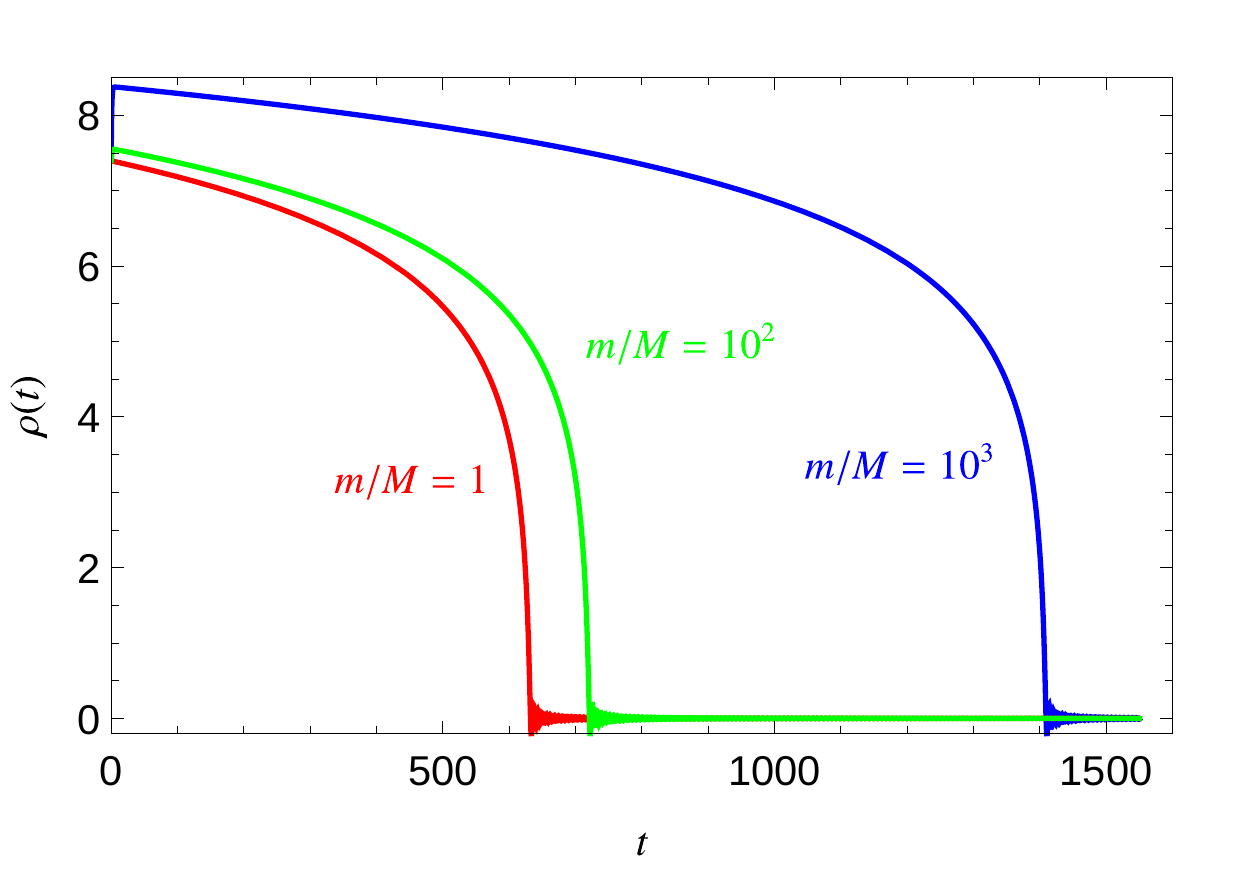}\caption{Left: The field $ \phi(t) $ as a function of time for the values of the ratio $m/M=1, 10^2, 10^3 $ . Right: The field $ \rho(t) $ as a function of time for the values of the ratio $m/M=1, 10^2, 10^3 $.}\label{phirho}
\end{center}
\end{figure}

\begin{figure}[t]
\begin{center}
\includegraphics[width=7.4cm]{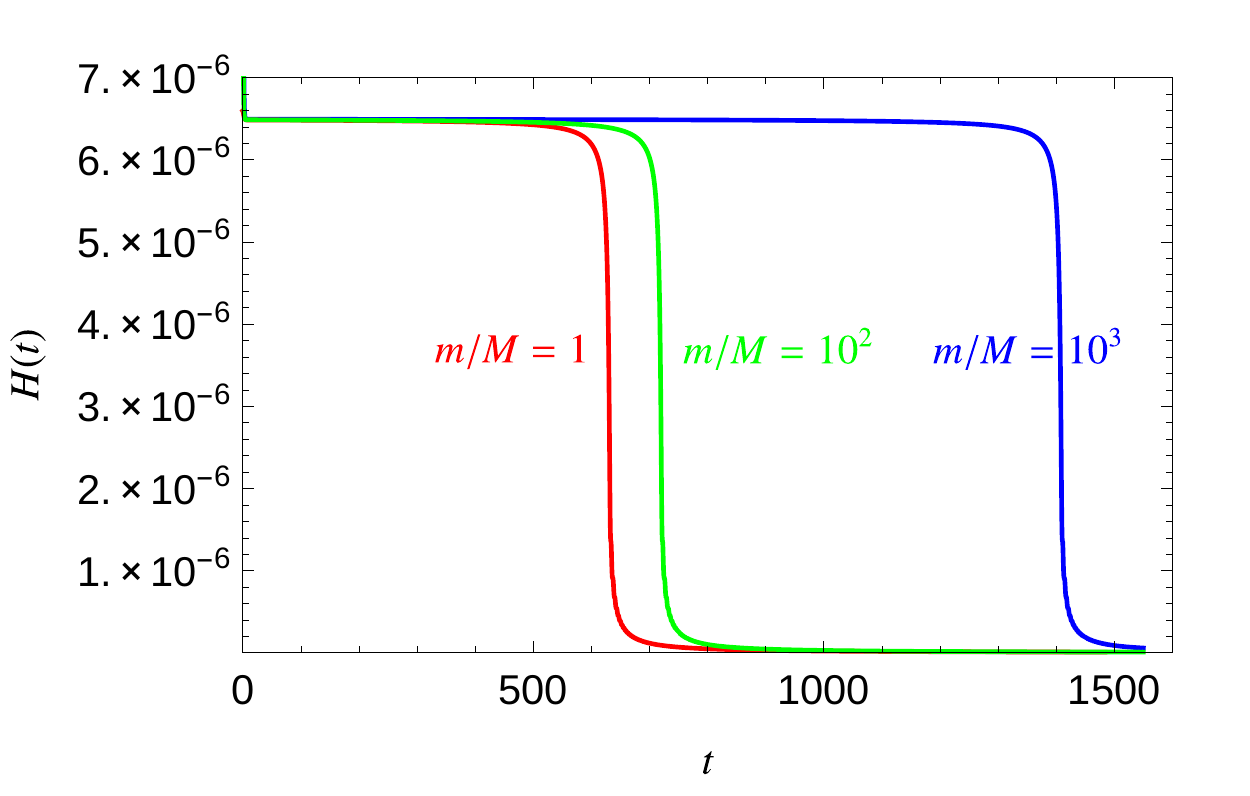}
\includegraphics[width=7cm]{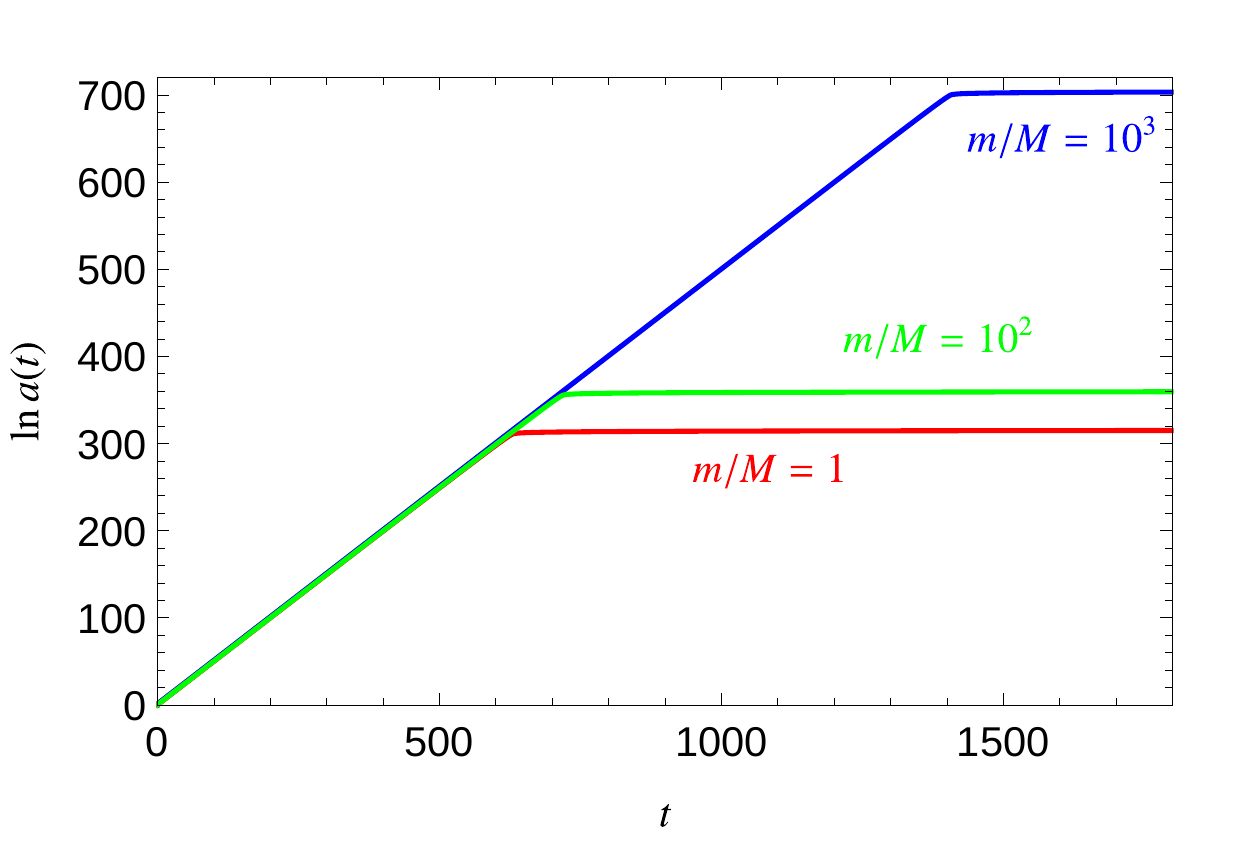}\caption{Left: The Hubble parameter as a function of time for the values of the ratio $m/M=1, 10^2, 10^3 $. Right: The logarithm of the scale factor $a(t)$ as a function of time for the values of the ratio $m/M=1, 10^2, 10^3 $.}\label{Hlna}
\end{center}
\end{figure}

In Figure \ref{pot} we see the profile of the potential for the choice of parameters $\alpha=0.01$, $\beta=0.001$ and $m=M=1.3\cdot 10^{-5}$. From this graph we observe, in contrast to the potential considered in the previous section, that in this potential we have a unique Minkowski vacuum, at the point $ (\phi_{\text{min}},\rho_{\text{min}})=(0,0). $  For the mass $m$ we study the cases between $m/M=1$ and $ m/M=10^3$. Our results for the fields, the Hubble parameter and the number of e-foldings, as functions of time for three typical values of the ratio $m/M$ are presented in Figure \ref{phirho} and Figure \ref{Hlna}. In all the cases we use the same initial conditions for the fields $\rho=7.4$ and $\phi=6$.

From these figures and our results we find that with increasing of the ratio $m/M$ we have new effects that are absent in the massless case. First, we observe that we have the damped oscillation of the field $\phi$, whose frequency is increased with increasing  the ratio $ m/M $. Thus we may have a possible contribution of the field $\phi$ to the reheating as $ m $ increases. This effect can be also observed for the massless case we studied in the previous section for larger values of the constant $\alpha$ (larger than 1), but for these values the viability of the inflationary model breaks down.  Second, we see that in the beginning we have a valid increase of the field $\rho$ which first reaches a maximum value and then starts the slow-roll inflationary process. This maximum value is increased with increasing the ratio $m/M$. Specifically, in the case $ m/M=1 $ the field $ \rho $ does not increase, in the case $ m/M=10^2 $ increases till the maximum value $ 7.6 $ and in the case $ m/M=10^3 $ till $ 8.4 $. Last, we notice that with increasing the ratio $m/M$ the duration of the inflationary period becomes longer.

From our results we find that the values for the observables in the case of including the mass term in the potential agree with the constraints \eqref{obs}, for values of $m/M$ ranging from $ 1 $ to $ 10^3 $, and thus also this case leads to viable inflation. For example for the case of $m/M=1$ the spectral index and the tensor-to-scalar ratio are $n_s=0.963\pm 0.004$ and $r= 0.0037 \mp 0.0007$, while in the case of $ m/M=10^3$, they are $ n_s=0.963\pm 0.003 $ and $ r= 0.0037 \mp 0.0007$. 


\section{The Case $U(\phi)=\frac{1}{4}\lambda (\phi^2-\upsilon^2)^2$}

In the following, we shortly study the case where $\phi$ is identified as the Standard Model Higgs boson. The scalar potential $V(\phi,\rho)$ in this case takes the form
\begin{equation}
\label{5.1}
V(\phi,\rho)= e^{-2\sqrt{\frac{2}{3}}\rho}\left[\frac{3}{4} \,M^2 (1+\beta \phi ^2) \left(1+\alpha \phi ^2 -e^{\sqrt{\frac{2}{3}}\rho}\right)^2+\frac{1}{4}\lambda (\phi^2-\upsilon^2)^2\right] \, .                                                                
\end{equation}
We focus on the case where $ \lambda=m_h^2/2\upsilon^2 \simeq 0.13 $ is fixed by the measured Higgs vacuum expectation value $ \upsilon \simeq 246\, GeV $ and Higgs boson mass $ m_h \simeq 125\, GeV $ at the electroweak scale. The case $ \beta=0 $ was exhaustively studied in \cite{Gundhi:2018wyz} where it is also being studied the case of smaller values of $ \lambda $ that are suggested by the Standard Model RG flow, which drives the running coupling $ \lambda $ to very small values at high energy scales. As discussed in \cite{Gundhi:2018wyz} and \cite{He:2018gyf} in the case  where $ \beta=0 $ the constraints in the parameter $ \alpha $ somehow relaxed, compared to the previous models, in order to be in agreement with all data. That is $ \alpha\lesssim 10^3-10^4\,. $ Such values for the parameter $ \alpha $ are also acceptable for $ \beta>0$.

The potential \eqref{5.1} is positive defined and it contains a two-fold degenerated Minkowski minimum and a saddle-point, being a minimum for $ \rho $ and a maximum for $ \phi $,  which respectively are given by
\begin{equation}
\label{5.2}
(\phi_{\text{min}},\rho_{\text{min}})=\left(\pm \upsilon , \sqrt{\frac{3}{2}}\ln(1+\alpha \upsilon^2) \right) \quad \text{and} \quad (\phi_{\text{sp}},\rho_{\text{sp}})=\left( 0, \sqrt{\frac{3}{2}} \ln(1+\frac{\lambda \upsilon^4}{3 M^2}) \right).
\end{equation}
At the extrema, the potential acquires the values
\begin{equation}
\label{5.3}
V(\phi_{\text{min}},\rho_{\text{min}})=0  \quad \text{and} \quad V(\phi_{\text{sp}},\rho_{\text{sp}})=\frac{\lambda}{4}\frac{1}{1/\upsilon^4+1/3M^2}.
\end{equation}
From the above equations we observe that the places of the extrema of the potential and the values of the potential calculated there do not depend on the constant $\beta$ and thus are the same with the places found for the case $\beta=0$ studied within \cite{Gundhi:2018wyz}. Furthermore, solving the equation $ \frac{d V}{d \phi}=0\, , $ it is trivial to find the trajectory that the inflaton takes place in the potential valley, but it is too complicated to be presented. Nevertheless in the limit where $ \beta\rightarrow 0 $ and for negligibly small $ \upsilon $ the known trajectory, $ \phi^2=\frac{e^{\sqrt{\frac{2}{3}}\rho}-1}{\alpha+\lambda/3\alpha M^2} $, is recovered. 

We study this model for the case where $ \alpha=0.01 $ and $ \beta=0.001 $ as in the previous sections and for different initial values for $ \phi $, $ \phi=6,\,1,\,0.6\,\, \text{and} \,\, 10^{-3} $ and $ \rho=7.4. $ The behaviour of the field $ \rho $ does not change in this range of initial values of $ \phi. $

In all cases the spectral index and the tensor-to-scalar ratio are $n_s=0.964\pm 0.004$ and $r= 0.0037 \mp 0.0007$, leading thus to viable inflation.

 
\section{Conclusion}

Inspired by the success of the Starobinsky model and the multifield nature of theories of particle physics we study the inflationary model proposed in \cite{Ketov}. In the begining, we review the feauture of a general  $F({\cal R},\phi)$ theory of gravity, which is conformally equivalent to an Einstein-Hilbert theory including two scalar fields. Specializing to the case $F({\cal R},\phi)= -\frac{1}{2} f(\phi) {\cal R} + \frac{1}{12M^2(\phi)}{\cal R}^2$ and $U(\phi)=0$, as considered in \cite{Ketov}, we found that it leads to a two-field potential differing from the one given in \cite{Ketov}. However  this also yields a viable inflationary model, as concluded therein.

Moreover, we considered other cases, as well, by adding a mass term for the field $\phi$ in the action. We found that this case also yields a viable inflationary model, while the evolution of the fields depends mostly on the ratio $m/M$. More specifically, we found that with increasing the ratio $m/M$ we have new effects that are absent in the massless case, such as the observation of damped oscillations of the field $\phi$, whose frequency is increased with increasing  the ratio $ m/M $. The ratio  $m/M$ affects  the duration of the inflationary period, too. Furthermore, in a Higgs-like potential, identifying the field $\phi$ with the Standard Model Higgs boson, we also found a viable inflation.

Finally, in the constraints we have provided for $\alpha$ and $\beta$ in the studied models we expect the non-gaussianities to be relatively small because $\phi$ is almost stabilized to its vacuum value before the last $50-60$ e-folds such that in these last e-folds the Starobinsky trajectory of $\rho$ drives the inflation. This observation also depends on the initial conditions for $\phi$ and $\rho$, but we checked that for initial conditions such that we have sufficient e-folds for inflation and the field $\phi$ contributes to the last $50-60$ of them, the predictions for $n_s$ were not in the range predicted by the observations.

Within the goals of future work, in the context of this model, and in order to check its viability, would be the study of reheating and preheating after the end of inflation, but this lies beyond the scope of this paper. Regarding possible extensions of this model these mainly include considerations of other physical scalar potentials, in the place of the $U(\phi)$ studied here and/or  the insertion of more complicated non-minimal couplings between gravity and the pre-existing scalar field which are motivated by higher dimensional theories, such as non-minimal kinetic terms and d' Alembertian of Ricci scalar terms, or Supergravity theories.


\acknowledgments

We are grateful to thank A.B. Lahanas, V.C. Spanos and N. Tetradis for useful discussions, comments and suggestions on the draft. D.C. thanks P. Christodoulidis for illuminating discussions. The research work was supported by the Hellenic Foundation for Research and Innovation (H.F.R.I.) under the “First Call for H.F.R.I. Research Projects to support Faculty members and Researchers and the procurement of high-cost research equipment grant” (Project Number: 824).


\end{document}